\documentclass[onecolumn,showpacs,preprintnumbers,amsmath,amssymb]{revtex4}

\usepackage{graphicx}
\usepackage{dcolumn}
\usepackage{bm}

\begin{document}


\title{Signatures of Many-Particle Correlations in Two-Dimensional 
Fourier-Transform Spectra of Semiconductor Nanostructures} 
 
\author{I.~Kuznetsova,$^{1}$  P.~Thomas,$^{1}$  T.~Meier,$^{1}$ T.~Zhang,$^{2,3}$ X.~Li,$^{2}$
R.~P.~Mirin,$^{4}$ and S.~T.~Cundiff$^{2}$}
\affiliation{$^{1}$Department of Physics and Material Sciences Center, 
Philipps University, Renthof 5, D-35032 Marburg, Germany}
\affiliation{$^{2}$JILA, University of Colorado and National Institute of Standards 
and Technology, Boulder, CO 80309-0440, USA}
\affiliation{$^{3}$Department of Physics, University of Colorado, Boulder, CO 80309-0390, USA}
\affiliation{$^{4}$National Institute of Standards and Technology, Boulder, CO 80305, USA}

\begin{abstract} 
On the basis of a microscopic theory, the
signatures of many-particle correlations in Two-Dimensional
Fourier-Transform Spectra (2D-FTS) of semiconductor nanostructures are
identified and compared to experimental data.
Spectra in the photon energy range of the heavy-hole
and light-hole excitonic resonances show characteristic
features due to correlations, which depend on the
relative polarization directions of the excitation pulses.
\end{abstract}

\maketitle
\label{Introduction} 


Recent reports illustrate the potential
of a novel method, known as ``Two-Dimensional Fourier-Transform Spectroscopy''
(2D-FTS), for the investigation of
many-particle induced correlations in semiconductor structures
\cite{steven05,borca05,steven06}.
2D-FTS is based on a four-wave-mixing experiment, where three
excitation pulses are separated in time and is heterodyne detected
to fully characterize its phase.
The signal is transformed into frequency domain
$\omega_t$ (emission frequency) and $\omega_{\tau}$
(excitation frequency), both
with respect to real time $t$ and time separation $\tau$
of the first two pulses,
respectively.
The third pulse is delayed with respect to the second one by $T$.
2D-FTS is widely used to study
vibrational \cite{ernst,Golonzka2001,Zanni2001} and electronic excitations
\cite{mukamel,Hybl2001,Brixner2005} in molecules.
Applying 2D-FTS in the optical regime we investigate the complex
interplay between exciton, biexciton and continuum excitations,
in semiconductor nanostructures.

In previous publications demonstrating 2D-FTS of
semiconductors \cite{steven05,borca05,steven06}, the
experimental results were compared to a phenomenological theory based
on extending the Optical Bloch Equations to include terms that describe
excitation induced dephasing (EID) \cite{Wang1993,Hu1994} and excitation
induced shift (EIS) \cite{Shacklette2002}. Here, we use a microscopic
theory to calculate 2D-FTS. Furthermore, the model predicts that
the 2D-FTS qualitatively depend on the polarizations directions of the excitation
pulses. Our results demonstrate
that for 2D-FTS, the influence of many-particle
correlations on the spectra can clearly be identified and are in
agreement with the experimental findings.
 
For the qualitative modeling of 2D-FTS 
in such systems and in order to keep the numerical  
requirement within reasonable limits  
we use a microscopic many-body theory and apply it to a 
one-dimensional tight-binding model.  
Therefore, a quantitative agreement between theory and experiment can  
not be expected. 
It has, however, been shown that many important signatures of nonlinear  
optical experiments 
performed on quantum wells can qualitatively well be reproduced 
by this model \cite{sieh_991,sieh_99,buch,meier00,brick01,finger02,euten99}.

The optical nonlinearities are treated up to third order in
the coherent $\chi^{(3)}$-limit beyond the Hartree-Fock level.
In order to separate the correlation effects
 from those due to the first-order Coulomb interaction (i.e.,
Hartree-Fock)
we use a set of equations for the interband coherence $P$
 and two-exciton amplitude $\bar B$ which in
symbolic form reads \cite{sieh_99,buch,axt,lindberg}
\begin{eqnarray}
\label{Modellpgleichung1}
-i\hbar \frac{\partial P}{\partial t} &=&
\hbar(\omega_X-i\gamma) P + V_C\  P^*\  P P - V_C
P^* \bar  B+ \mu^*E \nonumber\\
&-& \mu^*E\ P^*\ P,\\
\label{Modellpgleichung2}
-i\hbar  \frac{\partial \bar B}{\partial t} &=&
\hbar(\omega_{BX}+i\beta)  \bar B + P P {V}_C.
\end{eqnarray}
In fact, all terms in the microscopic version of the equations
depend on the spatial indices
of the tight-binding model as well as band indices, referring
to electron, heavy-hole (h) and light-hole (l) bands, respectively.
Details of the model and the set of microscopic equations 
that are used in our numerical evaluations are
 presented and discussed in Ref.\cite{buch}.
In Eqs.~(\ref{Modellpgleichung1}) and (\ref{Modellpgleichung2})
$V_C$ is the Coulomb matrix that is responsible
for the existence of exciton and biexciton resonances
 and interactions between them.
In Eq.~(\ref{Modellpgleichung1}) for the polarization $P$,
the Hartree-Fock terms are given by
the kinetic term
$\hbar \omega_X$, the renormalization term $V_C\  P^*\  P P$ and
the coupling to the light field containing the Pauli-blocking
$\mu^*E\ P^*\ P$ and the linear term $\mu^*E $.
The correlation term beyond Hartree-Fock is given by $V_C P^{*} \bar  B$.
As given in Eq.~(\ref{Modellpgleichung2}) the biexciton correlation $\bar B$ is driven by $PPV_C$.
The biexciton energy is $\hbar \omega_{BX}$. Coherent $\chi^{(3)}$-equations 
with h- and l-transitions have been previously analysed, see e.g., \cite{meier00,finger02,binder02}.

The quantum well samples used in this study show
heavy hole and light hole excitonic resonances.
Their spectral positions are determined by
the kinetic parts of Eqs.~(\ref{Modellpgleichung1}) and (\ref{Modellpgleichung2}),
 which also contain the phenomenological
dephasing rates,
$\gamma=1/T^{(h,l)}_2$ for excitons
and  $\beta=1/T^{(hh,ll,hl,lh)}_2$ for biexcitons.
 By such rates we can distinguish biexcitons formed by h-excitons ($hh$),
 by l-excitons ($ll$) and mixed h- and l-excitons ($hl, lh$).
 The optical transitions are given by
 dipole matrix elements $\mu$ chosen to model the
relevant selection rules for dipole transitions in III-V
semiconductor structures
\cite{buch,brick01}.
 This model agrees with the
selection rules for the nonlinear polarization
presented by M. Lindberg et al. for the three beam situation
\cite{lindberg94}.

Based on this model we
present  microscopic calculations
for 2D-FTS for semiconductor structures. 
We choose realistic  material parameters,
 ratio between h- and l-effective masses  and dipole
matrix elements, energetic h- and l-offset,
and Coulomb strength.
The experimental
conditions such as pulse duration, detuning, dephasing times,
polarization directions and the sequence of the pulses
are incorporated into the model. The dephasing times of
the various excitons and biexcitons are taken as
phenomenological parameters.

In this paper,
we concentrate on the situation where the first
pulse enters the nonlinear polarization conjugated, the so called
rephasing case \cite{steven06}.
The figures show
the amplitude $|E(\omega_{t},\omega_{\tau},T)|$, where
the electric field $E(\omega_{t},\omega_{\tau},T)$ is proportional to the
third-order polarization multiplied by the imaginary unit, i.e.,
$iP(\omega_{t},\omega_{\tau},T)$.
The emission frequency, $\omega_{t}$, is given on the horizontal-axis. The emission
frequency is used to define the sign of frequency, thus the conjugation of
the first pulse means that the
 excitation frequency, $\omega_{\tau}$, is negative. It is given on the vertical
axis.
The h- and l-exciton peaks then show up on the diagonal in the
left-upper (low energy)
and right-lower (high energy)
corners, respectively, while the off-diagonal peaks are due to
coupling between h- and l-exciton resonances.
 
\label{experiment} 

For the particular system considered in this paper, i.e., a GaAs quantum well, the
nonlinear optical properties have been explored quite extensively
in the past using four-wave-mixing and pump-probe experiments. This
system therefore
presents a good testing ground for the exploration of the
features of 2D-FTS. In particular, the sample used in the experiment is
a 10 period multiple quantum well structure with 10 nm GaAs wells
and 10 nm AlGaAs barriers. It is held at 8K. Experimental spectra
are obtained using the apparatus described by Zhang et al. \cite{steven05},
with the addition of polarizing optics.
 
\label{Results_discussions}

In the following we 
concentrate on 
the amplitude features of 2D-FTS. 
The distribution of the heights of the peaks
strongly depends on the interplay of 
the material and experimental parameters, 
including tuning of the excitation pulses, shape and 
temporal full width at half maximum (FWHM) of the pulses, ratio  
$\mu^{h}/\mu^{l}$ of h- vs. l-dipole 
matrix elements (which due to band-mixing in the quantum well structure 
depart from their bulk values) and dephasing times. 
We found that $|\mu^{h}/\mu^{l}|^2=2.1$ is a good choice for modeling the 
experiment.   
The simplest approach to model the   
h-l-exciton system is a three level system (V-system)  
without any 
interactions, but with coupling of two excited levels (modeling the h- and  
l-excitonic resonances)  
to a common ground state. In this model  
it is not possible to obtain  
a peak height distribution  
in the 2D-FTS features that is asymmetrical with respect to the diagonal, 
if the excitation pulses are extremely short and $\mu^{h} = \mu^{l}$.  
The asymmetry of such spectra in our experiments on quantum well systems  
appears because of the interplay of several effects:  
different dipole matrix elements, 
dephasing times of excitons and biexcitons, degree of 
overlap with the spectrum of 
the excitation 
pulses and many-particle couplings within the system,  
as supported by the theoretical  
spectra. In our model the electric field 
of the excitation pulses is proportional to $\exp{(-t^2/\delta^2)}$, where 
$\delta$ is the Gaussian width of around 115 fs, corresponding to  
intensity FWHM of 135 fs and  
a spectral FWHM  
of 13.5 meV. 
   
\begin{figure}[t!] 
\begin{center} 
\includegraphics*[width=16cm]{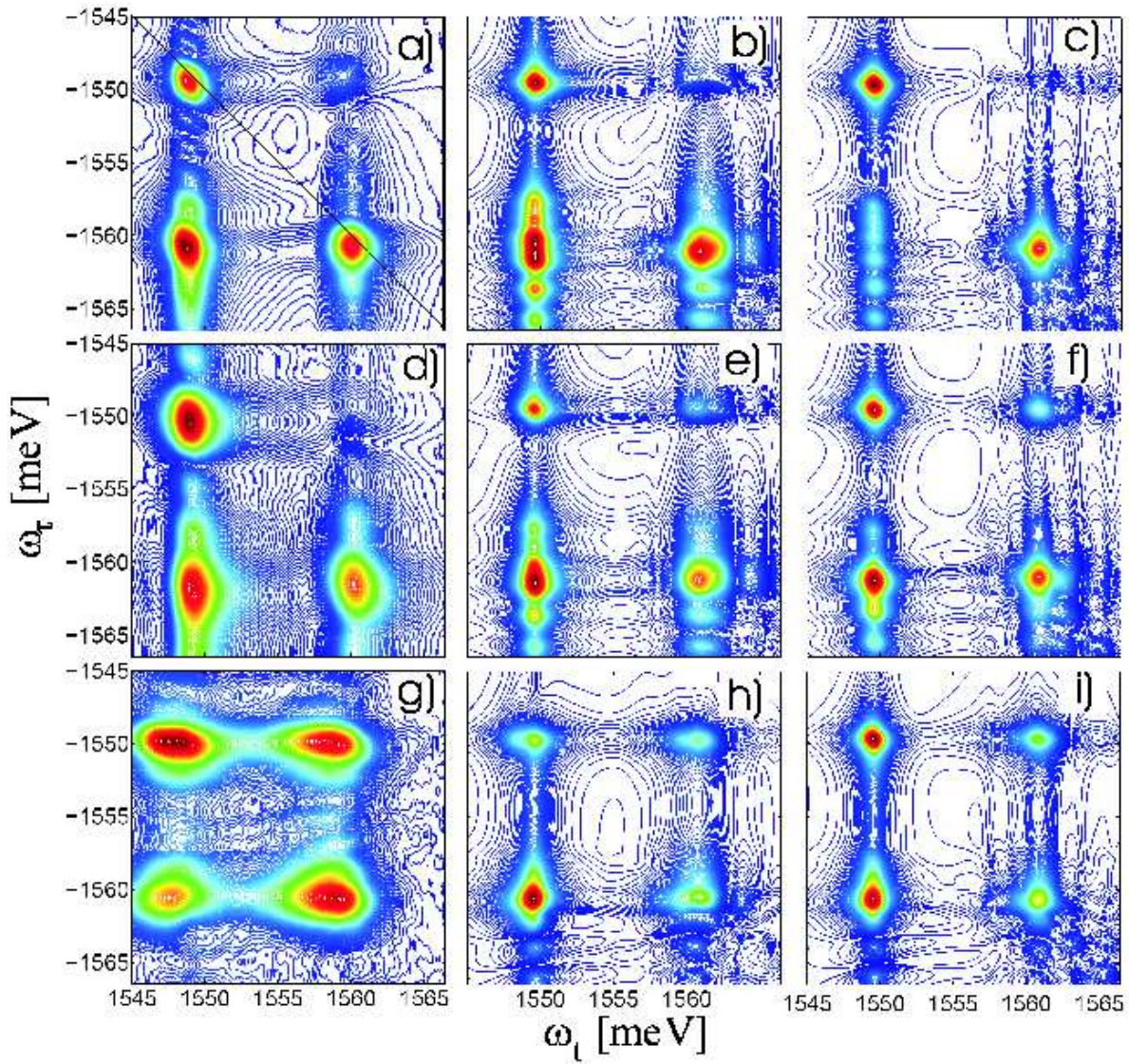} 
\end{center} 
\caption{2D-FTS spectra amplitudes (normalized). Left column: experimental. 
Middle column: full $\chi^{(3)}$-calculation. Right column:  
Hartree-Fock calculation. Upper row: $\sigma^+\sigma^+\sigma^+$ polarization,
$T=100$fs. 
Middle row: $XXX$ co-linear polarization,
$T=100$fs. Lower row: $YXX$ cross-linear 
polarization, $T=350$fs. 
Dephasing times are: $T_2^{h}=1.3$ps, $T_2^{l}=0.8$ps 
for excitons  
and $T_2^{hh}=1.3$ps, $T_2^{ll}=0.8$ps, $T_2^{hl}=T_2^{lh}=0.5$ps 
for biexcitons. Theoretical spectra are weakly
inhomogeneously broadened by Gaussian width of 
0.3 meV.} 
\label{exp} 
\end{figure}

In Figure \ref{exp} 
we compare experimental (first column) and theoretical results for the 
amplitudes from the full calculation (second 
column) and the Hartree-Fock part (third column)  
for $\sigma^+\sigma^+\sigma^+$, co-linear $XXX$, and cross-linear 
$YXX$  
polarizations. 
In all figures   
the central excitation energy is 1meV above the  
l-exciton in order to compensate for the small 
dipole matrix element of the l-exciton. 
 
Numerical results for the full calculation (second column) show agreement with experiments,
not only for the h-, l-exciton and mixed peaks,
but also for the h-continuum at higher
$\omega_{\tau}$ (the vertical continuum contributions in the experimental data
are more apparent in a previous publication \cite{borca05}).
Note that in the theoretical figures, the
h-continuum, appearing close to the l-excitonic
peaks as dominantly vertical structures,
is decomposed into discrete
peaks because only $N=40$
sites have been taken in the tight-binding approach for
numerical reasons. It should be noted that the continuum appears
dominantly as a vertical feature close to the mixed peaks
in the lower-left corner. This demonstrates
that the lower energetic part of the
continuum can not be seen as a pure inhomogeneous spectrum in the
model containing the many-particle interaction
\cite{buch,irina}. In an interaction-free 
model the continuum would in fact be purely inhomogeneous, i.e., it  
would consist  
of independent two-level systems belonging to a spectrum of $\vec k$-vectors. 
If this were true, one would expect a feature of 
continua along the main diagonal close to 
the l-excitonic peak in the right-lower corner. In 
the frame of the Hartree-Fock approximation, where the correlations are 
ignored, the features due to the h-continuum clearly show up 
at the higher energies for all polarization cases 
(Fig. \ref{exp} third column).

The co-linear situation (Fig. \ref{exp} middle row) 
looks similar to the co-circular case (Fig. \ref{exp} upper row), as 
is to be expected from the selection rules.  It is remarkable, that  
the upper right non-diagonal mixed peak for co-circular case 
(Fig. \ref{exp}b) is very small. As a 
comparison with a Hartree-Fock result (Fig. \ref{exp}c) shows, this 
is not due to 
correlations, but results from an interplay 
of detuning, dipole matrix elements, and dephasing times.  
The mixed signature in the left-lower corner for the Hartree-Fock calculation 
(Fig. \ref{exp}c) 
 shows discrete weak features due to the h-l-excitons
and the h-continuum. 
 The many-body correlations lead to the increase of this mixed signature 
at higher excitation energy 
 and vertical features of the continuum as well. Unlike to the 
co-circular case,  
the influence 
of correlations on the
mixed signature at higher excitation energy is less pronounced for the
co-linear case. The
Hartree-Fock approximation 
(Fig. \ref{exp}f) and the full 
calculation 
(Fig. \ref{exp}e)  
do not differ that much for the co-linear case. 
This can be understood  by consideration 
of a  
simple V-system, where for 
the excitation $\sigma^+\sigma^+\sigma^+$ the h-  
and l-transitions resemble
two independent two-level systems. 
 The linear case contains both $\sigma^+$ and $\sigma^-$ transitions,
thereby such 
excitation  
couples the h- and l-exciton transitions resulting in two V-systems.  
This coupling generates the mixed signatures 
and appears for both linear polarized cases  
in the Hartree-Fock calculations (Fig. \ref{exp}f: XXX, i: YXX).

The cross-linear polarized excitation case ($YXX$)  
shows additional signatures from  
two-exciton resonances (bound and unbound biexcitons), as predicted by 
the selection 
rules and supported by the theoretical results (Fig. \ref{exp}h). This is due to  
a suppression of the 
excitonic features with respect to the biexcitonic ones \cite{finger02}.  
Bound biexcitons show up on the low emission-energy side and  
unbound ones on the high emission-energy side 
of the exciton \cite{sieh_991,sieh_99,finger02}. The biexciton contribution results in a
 horizontal elongation of the peaks compared to the $XXX$-case. 
 The Hartree-Fock calculation (Fig. \ref{exp}i) 
clearly shows, that the horizontal elongation of the $YXX$ spectra (Fig. \ref{exp}h) 
is due to correlations, i.e., bound and unbound two-exciton states. In the 
Hartree-Fock limit the horizontal elongation is absent (Fig. \ref{exp}i).  
Interestingly, the continuum 
contribution and respectively the elongation of the signatures at higher 
excitation  
energy is less developed as compared to 
the co-linear case, which is supported by the experiment (Fig. \ref{exp}d,g). 
For the cross-linear situation one can show \cite{buch,irina} that 
indeed the continuum becomes suppressed due to cancellation effects 
that result from Coulomb-correlations in this situation. Indeed, the 
Hartree-Fock calculation shows a more pronounced continuum  
contribution (and concomitantly of  
the superimposed l-exciton peak)  
in the right-lower corner of the spectrum, as compared to the 
full $\chi^{(3)}$-calculation.

It is surprising that the lower left non-diagonal peak is  
the strongest one in the theoretical spectra for the 
$YXX$-case (Fig. \ref{exp}h),  
in contrast to the experimental data (Fig. \ref{exp}g), 
where the h-peak is the strongest one. 
At present, the reason for this discrepancy is not clear.   
The peak in question is a mixture of three contributions, the 
h- and l-excitonic resonances and the h-continuum. Therefore 
it is reasonable to 
assume an effective dephasing of this peak, which is not  
included in our present treatment. 

A comparison between the results of the full and the Hartree-Fock 
calculation shows that the left-lower non-diagonal peak  
is less prominent in the latter spectra. This difference suggests that 
correlations enhance this signature in the full treatment. These 
couplings could well be more sensitive to relaxation  
processes than the trivial ones which already appear in the V-system 
or within a model, which only considers 
the phase-space (Pauli-blocking) nonlinearities. 
 
We have not studied the dependence of the peak heights
on the spectral shape of the incident laser pulses which
we have here always taken as Gaussian. Our calculations
have shown that the temporal duration and thus the spectral
width of the laser beams strongly influences the 2D-FTS.
Therefore, deviations of the pulse envelopes from a
Gaussian shape will clearly significantly alter the
peak heights.

The amplitude of the non-diagonal signatures relative to the 
diagonal ones generally depends on  $T$.  
Already for a V-system one finds that the  
two non-diagonal peaks show beating with respect to $T$, where 
the period is 
$2\pi\hbar/\Delta E_{h,l}$, with $\Delta E_{h,l}$ the energetic 
h- and l-exciton separation. In the experimental data 
shown here $T=350$fs has been taken. With $\Delta E_{h,l} 
= 11.5$meV, corresponding to a beating period of  
360 fs, this is close to the situation of maximum non-diagonal peaks. 
However, as mentioned above, the left-lower signature is  
definitely too strong if compared to the experiments.  
In our calculations for the $YXX$-case we have used $T=350$ fs. 
 
It should be noted that 
biexcitonic features can be observed in different kinds of  
nonlinear optical experiments 
for certain polarization cases. The 2D-FTS always show the 
presence of biexcitonic features in the plots for amplitude   
or real part (not shown here), 
albeit to different extent, depending on polarization.    

Since 2D-FTS is based on a photon-echo experiment, it is not surprising
that it can be used to measure the dephasing time $T_2$ even in the presence
of inhomogeneous broadening. Specifically, inhomogeneous broadening will
elongate peaks along the diagonal, while the width perpendicular to the
diagonal is inversely proportional to $T_2$. However, 2D-FTS is clearly
superior to a traditional one-dimensional photon echo measurement when
multiple resonances are involved, as is the case here. The photon-echo
signal mixes the contributions of the resonances, possibly leading to
ambiguity in assigning the dephasing times. This ambiguity does not
occur in 2D-FTS as the peaks corresponding to distinct resonances
are well separated.

A traditional one-dimensional transient four-wave-mixing (or photon
echo) measurement also has trouble yielding $T_2$ when there are
strong many-body correlations, which may result in a rapid decay
of the signal unrelated to $T_2$ \cite{cundiff1996}. 2D-FTS
is much better at revealing $T_2$ of individual resonances. 
As an example we consider theoretical 2D FTS-amplitude plots 
for cross-linearly polarized pulses, which show the 
correlation-induced features most prominently.
\begin{figure}[h!] 
\begin{center} 
\includegraphics*[height=7cm]{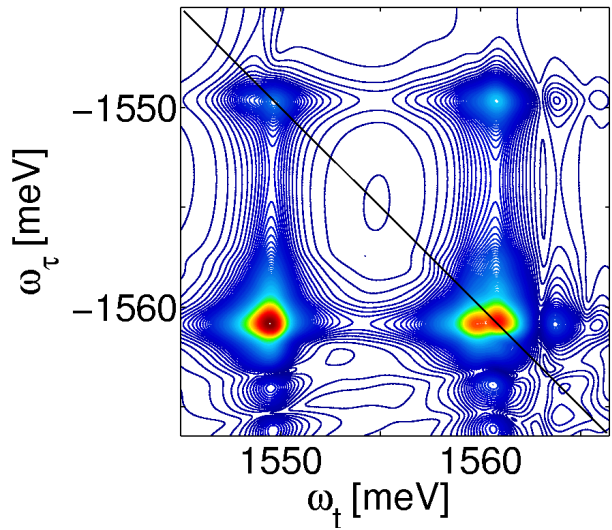} 
\end{center} 
\caption{Amplitude of 2D-FTS for YXX case. 
 Dephasing times are: $T_2^{l}=T_2^{h}=1.3$ps for excitons  
and $T_2^{ll}=T_2^{hh}=1.3$ps, $T_2^{hl}=T_2^{lh}=0.65$ps 
for biexcitons. 
Theoretical spectra are slightly
inhomogeneously broadened by Gaussian width of 
0.3 meV.}
\label{Texc} 
\end{figure} 
However, now the l-transitions 
have been given the same dephasing time as the h-ones, 
i.e., $T_2^{l}=T_2^{h}=1.3$ps, while the theoretical  
results shown in Figure \ref{exp} correspond to 
$T_2^{h}=1.3$ps and $T_2^{l}=0.8$ps. Figure \ref{Texc} shows, that 
 the distribution of the peak heights changes due to longer dephasing time of l-exciton.  
The comparison with Fig. \ref{exp}h demonstrates the 
large sensitivity of the 2D-spectra on $T_2$-times.  
 
\label{Conclusions}

In summary, we have investigated correlation effects
 in quantum well  systems by  applying the Two-Dimensional
Fourier-Transform Spectroscopy (2D FTS). This method provides a wide spectrum of 
information about the many-body correlations simultaneously, such as:
the strength of the couplings between excitons, biexcitons and continua,  
character of continuum excitations, 
dephasing times, and dependences of 
these features on the polarization directions.

A first comparison  
between experiment and theory has been performed 
on the basis of a microscopic model. Depending 
on polarization directions of the excitation pulses, characteristic 
signatures of many-particle correlations can be identified in the 
amplitude spectra. Differences between experiment and theory point towards 
the action of relaxation processes on many-particle 
correlations, which are not included in the present 
purely coherent treatment.  Additional important information is contained in the 
real- and imaginary part of the spectra. Work is in progress to 
analyse these parts of the signal and to
exploit these spectra for 
semiconductor structures in the energetic range 
of both excitonic and continuum transitions.

 
\label{acknow} 
This work has been supported by the Optodynamics Center of
the Philipps-University Marburg,
and by the Deutsche Forschungsgemeinschaft (DFG),
and by the
Research Centre J\"ulich (NIC).
T.M. thanks the DFG for support via a Heisenberg fellowship (ME
1916/1). The work at JILA was supported DOE/BES grant DE-FG02-02ER15346.


\end{document}